\documentclass[twoside]{article}
\usepackage{fleqn,espcrc2}

\usepackage{graphicx}

\newcommand{\AmS}{{\protect\the\textfont2
  A\kern-.1667em\lower.5ex\hbox{M}\kern-.125emS}}

\title{Signatures of Spin Degrees of Freedom in Charge Dynamics of
  YBa$_2$Cu$_3$O$_{6.95}$}

\author{E. Schachinger\address{Institut f\"ur Theoretische Physik,
        Technische Universit\"at Graz, A-8010 Graz, Austria}
        and 
        J.P. Carbotte\address{Department of Physics and Astronomy, McMaster
        University, Hamilton, Ont. L8S 4M1, Canada}}
       
\begin{document}

\begin{abstract}
Charge carrier-spin excitation spectral densities constructed from
optical data and varying with temperature are used to calculate
various properties of the charge carrier dynamics. Temperature
$(T)$ variations of such properties are sensitive to the changes
in the spectral density brought about by the growth in the
$41\,$meV spin resonance as $T$ is lowered. The microwave
conductivity and the thermal conductivity are stressed and
comparison with experiment is given. The imaginary part of the
conductivity as a function of frequency $\omega$ is also
featured.
\vspace{1pc}
\end{abstract}

\maketitle

Following the demonstration by Carbotte {\it et al.} \cite{Carb1}
that infrared optical conductivity data
could be used to identify the strength and form of the coupling
between the charge carriers and the spin degrees of freedom in
optimally doped YBa$_2$Cu$_3$O$_{6.95}$ (YBCO), Schachinger and
Carbotte \cite{Schach2} extended the analysis to other high
$T_c$ materials. They found that the existence of a spin resonance
and its coupling to the charge carriers is not limited to the
bilayered materials although it does not exist in all high $T_c$
superconductors. The essence of the analysis whereby a spectral
density is obtained, involves a second derivative of the optical
scattering rate. This was first introduced by Marsiglio {\it et al.}
\cite{Mars1} for the electron-phonon case in the normal state,
and applied with considerable success to the case of K$_3$C$_{60}$,
a narrow band system in which correlation effects are expected to
be significant. The method was generalized by Carbotte {\it et al.}
\cite{Carb1} to the superconducting state with $d$-wave order
parameter. This involves generalized Eliashberg equations in
which the dominant coupling is assumed to be due to the spins
as in the nearly antiferromagnetic Fermi liquid theory by
Pines and coworkers \cite{Millis1,Mont1}. The interactions enter
only through the carrier-spin spectral function $I^2\chi(\omega)$.
Further work by Schachinger {\it et al.} \cite{Schach3} showed
that optical data at various temperatures could be used to obtain
a spectral density at each temperature and that these functions
reflect the growth of the $41\,$meV spin resonance measured by
inelastic spin polarized neutron scattering \cite{Dai}. A temperature
dependence to the  effective spectral density $I^2\chi(\omega)$ is different
from the case of the electron-phonon interaction where there is
no significant $T$ dependence. It
is characteristic of highly correlated electron systems in which the
effective interaction responsible for superconductivity resides in
the electron system itself and therefore is modified by any change
in the state of the charge carriers. Spin fluctuations in the
nearly antiferromagnetic Fermi liquid model are an example of such a
mechanism but is not unique.

In the work of Schachinger {\it et al.} \cite{Schach3} the temperature
dependence of the spectral weight under the $41\,$meV resonance which
contributes to the pairing interaction agrees well with the neutron
data Dai {\it et al.} \cite{Dai} on the growth of the resonance peak
as $T$ is lowered in the superconducting state. The resulting spectra
$I^2\chi(\omega)$ lead to good agreement with experiment for the
temperature dependence of the in-plane superfluid density as well as
\begin{figure}[t]
\vspace*{-12mm}
\hspace*{-3mm}
\includegraphics[width=71mm]{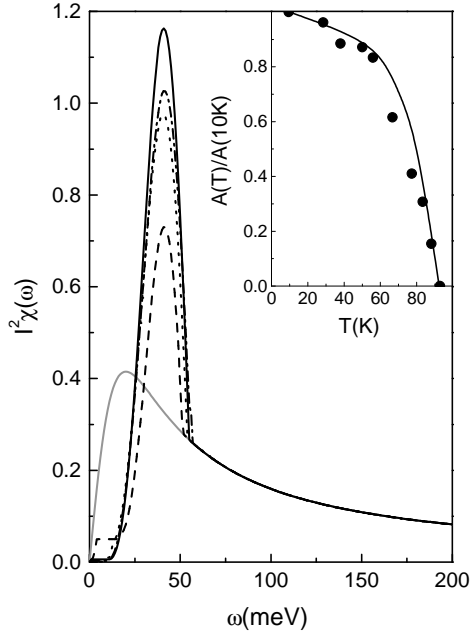}
\vspace*{-12mm}
\caption{The charge carrier-spin excitation spectral density
$I^2\chi(\omega)$ determined from optical scattering rate data at
various temperatures. Gray curve $T=95\,$K, dashed $T=80\,$K,
dotted $T=60\,$K, dash-dotted $T=40\,$K, and solid $T=10\,$K. Note the
growth of the strength of the $41\,$meV spin resonance as the
temperature is lowered. The insert gives the spectral weight
under the resonance $A(T)$ normalized to its $T=10\,$K value
(solid curve) compared with the neutron data of Dai {\it et al.}
\cite{Dai} for the equivalent quantity (solid circles).}
\label{f1}
\end{figure}
for the large peak observed well below $T_c$ in the microwave
conductivity. This peak reflects the collapse of the inelastic
scattering rate in the superconducting state due to the gapping of
the low energy part of the spectral density by the onset of
superconductivity. The calculations also offer an explanation for the
large value of the gap to critical temperature ratio observed as
compared to the BCS $d$-wave value. They give the correct zero
temperature condensation energy per copper atom and they predict
correctly that the superfluid density is only about 25\% of the
total optical oscillator strength. This arises because only the
coherent quasiparticle part of the electronic spectral function
condenses into the superfluid. The large incoherent background remains
largely unaffected.

Here we extend the work to include the temperature dependence of the
thermal conductivity and discuss in some detail the fit obtained
to optical conductivity data with emphasis on
the signature of the spin resonance in this
quantity.

In Fig.~\ref{f1} we show the charge carrier-spin excitation spectral
density $I^2\chi(\omega)$ obtained in the work of Schachinger
{\it et al.} \cite{Schach3} at various temperatures, namely
$T=95\,$K (gray), $T=80\,$K (dashed), $T=60\,$K (dotted),
$T=40\,$K (dash-dotted), and $T=10\,$K (solid). These are constructed by
fitting a spin fluctuation form for the effective charge carrier-%
spin excitation spectral density $I^2\chi(\omega)$ to optical
data via generalized $d$-wave Eliashberg equations. First a
two parameter form (MMP \cite{Millis1}) is fit to the normal state
at $T=95\,$K,
$I^2\chi(\omega) = I^2\omega/(\omega^2+\omega^2_{SF})$, where
$I$ is a measure of the charge-spin coupling and $\omega_{SF}$
is an energy characteristic of the spin fluctuations. The fit to the
$T=95\,$K optical scattering rate, $\tau^{-1}(\omega)$, determines
both parameters. $I^2$ is adjusted to get the
measured magnitude of $\tau^{-1}(\omega)$ and $\omega_{SF}$ to
reproduce its observed frequency dependence. As the temperature is
lowered into the superconducting state the optical data shows the
growth of a new collective mode at $41\,$meV in YBCO. The second
derivative of $\omega\tau^{-1}(\omega)$ with respect to $\omega$
gives a first estimate of the position, shape, and size of this
\begin{figure}[t]
\vspace*{20mm}
\hspace*{-3mm}
\includegraphics[width=85mm]{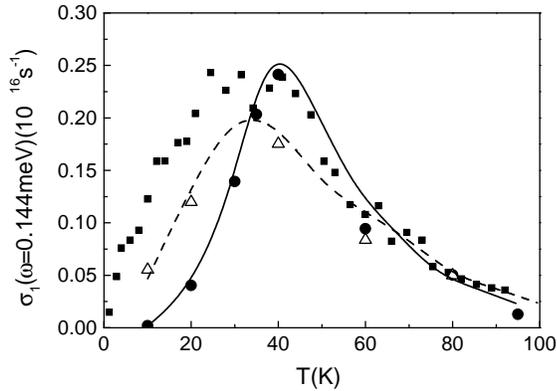}
\vspace*{-37mm}
\caption{The microwave conductivity at
$0.144\,$meV as a function of temperature $T$.
The solid squares are the data of Bonn {\it et al.} \cite{Bonn1},
the solid circles our theoretical results
for the pure case, while
the open triangles include some impurity scattering
$\tau^{-1}_{imp}\simeq 1.8\,$meV. The solid and dashed curves are
guides to the eye.}
\label{f2a}
\end{figure}
contribution to the spectral density which is then used to
modify $I^2\chi(\omega)$ in this energy range.
Other frequency regions are left
unaltered. The curves for $I^2\chi(\omega)$ change very significantly
with $T$ (see Fig.~\ref{f1}).
Also, as shown in the inset of Fig.~\ref{f1},
the spectral weight under the resonant peak agrees well
with the spin polarized neutron scattering data of Dai {\it et al.}
\cite{Dai} for the equivalent quantity.

Once the spectral density of Fig.~\ref{f1} is known at a particular
temperature, the $d$-wave Eliashberg equations can be solved for
the frequency dependent gap and renormalization function and
various properties can be determined from them. As a first result
we show in Fig.~\ref{f2a} the temperature dependence
\begin{figure}[t]
\vspace*{20mm}
\hspace*{-3mm}
\includegraphics[width=85mm]{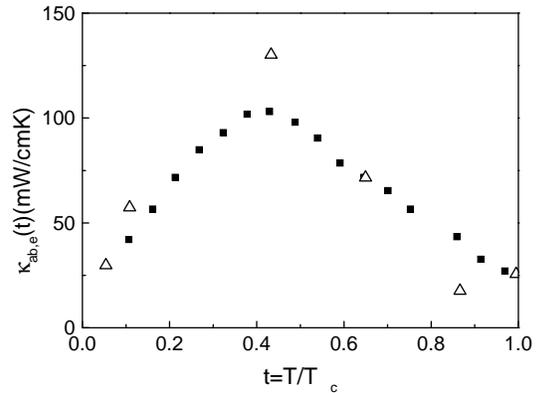}
\vspace*{-37mm}
\caption{The electronic part
of the thermal conductivity as a function of temperature $T$. The
solid squares are the experimental results of Matzukawa {\it et al.}
\cite{Matsuk} and the open triangles our theoretical result.}
\label{f2b}
\end{figure}
of the microwave peak at $\omega=0.144\,$meV \cite{Schach3}. The
solid squares are the experimental data of Bonn {\it et al.} \cite{Bonn1}.
The solid circles are our theoretical results assuming a clean sample,
i.e. no impurity scattering. We see that there is good agreement at
higher temperatures but that the theoretical curve is not broad
enough as $T$ is reduced. Much better agreement results when a small
amount of impurity scattering is included with
$\tau^{-1}_{imp}\simeq 1.8\,$meV (open triangles). The solid and dashed
curves are from previous work \cite{Schach4} based on a phenomenological
spectral density but can also be considered as guides to the eye.

A second quantity of interest is the temperature dependent electronic
thermal conductivity $\kappa_{ab,e}(T)$ (in-plane). This is shown
in Fig.~\ref{f2b}. The solid squares are the
experimental results of Matsukawa {\it et al.} \cite{Matsuk} and the
open triangles are our numerical results.
We see that the agreement is
good and that our spectral densities are capable of describing the
observed transport properties of optimally doped YBCO.

\begin{figure}[t]
\vspace*{20mm}
\hspace*{-3mm}
\includegraphics[width=85mm]{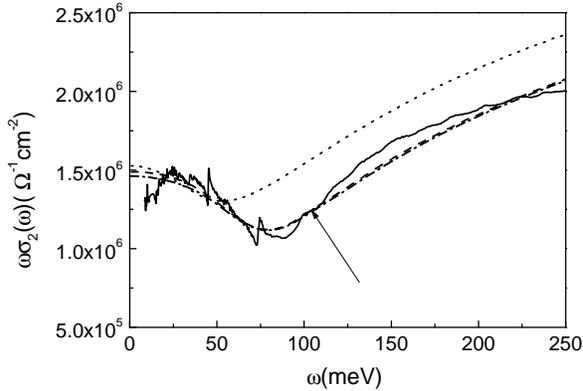}
\vspace*{-37mm}
\caption{The imaginary part of the optical conductivity,
$\sigma_2(\omega)$, times $\omega$ as a function of $\omega$. The
solid black curve is the data \cite{Homes}, the dotted curve theoretical
results based on the MMP spectral density with $\omega_{SF} = 20\,$meV
without modification for the $41\,$meV resonance peak while the dashed
and dash-dotted curve include it. Dashed is for the pure case,
dash-dotted includes
elastic impurity scattering with $\tau^{-1}_{imp}\simeq 1.8\,$meV.
The arrow indicates the data point used to fit theory to experiment.}
\label{f3a}
\end{figure}

In the procedure followed to obtain the $I^2\chi(\omega)$ of Fig.~\ref{f1}
only the optical scattering rate $\tau^{-1}(T,\omega) = \Omega^2_p
\{\Re{\rm e}[\sigma^{-1}(T,\omega)]\}/4\pi$ is used. Here
$\sigma(T,\omega)$ is the infrared a.c.\ conductivity as a function of
frequency $\omega$, at temperature $T$, and $\Omega^2_p$ is the plasma
frequency obtained from the in-plane optical oscillator strength. In
Fig.~\ref{f3a} we show our theoretical results for $\omega$
times the imaginary part of the optical conductivity,
$\omega\sigma_2(\omega)$, as a function of $\omega$ at $T=10\,$K and
compare with data of Homes {\it et al.} \cite{Homes} (solid curve).
Several points can be made. First, as shown by
Schachinger {\it et al.} \cite{Schach3} the limit $\omega\to 0$
of $\omega\sigma_2(\omega)$ fits well the measured temperature
dependence of the superfluid density. Second, the dotted curve which
is included for comparison was obtained using the normal state
MMP \cite{Millis1} spectrum
and does not include the $41\,$meV resonance. The other two curves
in Fig.~\ref{f3a} (dashed, pure case, dash-dotted with some
impurities) do and this has improved the agreement
with the data substantially. The minimum at $\omega=75$ to
$80\,$meV (equal to approximately twice the gap plus the resonance
frequency $\omega_{sr}$) can be traced to the resonance peak
and is its signature. No such
minimum would exist in a BCS formulation of $d$-wave superconductivity.

\begin{figure}
\vspace*{20mm}
\hspace*{-3mm}
\includegraphics[width=85mm]{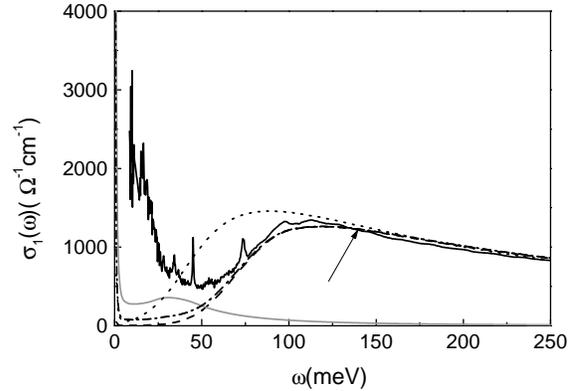}
\vspace*{-37mm}
\caption{The real part, $\sigma_1(\omega)$, of the
optical conductivity with various curves the same as in Fig.~\ref{f3a}.
The additional solid gray curve is the result of a BCS
$d$-wave calculation.}
\label{f3b}
\end{figure}

In Fig.~\ref{f3b} we show results for the real part
of $\sigma(T,\omega)$ at $T=10\,$K.
The solid gray curve is BCS and shows no agreement with
experimental data (black curve). It is clear that the high $T_c$
superconductors are far from simple BCS superconductors modified
to account for the $d$-wave symmetry of the gap. The dotted curve
gives results of Eliashberg calculations based on the solid gray
curve of Fig.~\ref{f1} for $I^2\chi(\omega)$ which does not include
effects of the $41\,$meV spin resonance. This curve is important
for comparison with results of a calculation which includes the
resonance, dashed curve in the clean limit and dash-dotted curve with a
small amount of elastic impurity scattering. Including
the $41\,$meV resonance in the spectral density
has improved agreement with experiment by shifting the main
rise in the conductivity (after the remaining Drude like part at very
small $\omega$ has largely died out) to higher energies.

In conclusion, the formation of the $41\,$meV spin resonance in the
superconducting state of optimally doped YBCO can be tracked accurately
through the optical conductivity which carries a signature of its growth
with decreasing temperature. The resonance plays an important role
in the transport and thermodynamic properties of YBCO leading to
distinct temperature dependences and characteristic structure in
spectroscopic quantities as a function of frequency $\omega$.

Research supported in part by NSERC (Natural Sciences and
Engineering Research Council of Canada) and by CIAR (Canadian
Institute for Advanced Research). The authors are grateful to
Dr. D.N.~Basov and Dr. C.C.~Homes for supplying their original experimental
data.

\end{document}